\documentclass[aps,prd,showpacs,twocolumn,nofootinbib]{revtex4}
\usepackage{graphicx}
\usepackage{dcolumn}
\usepackage{bm}
\usepackage{float}

\begin{document}

\def\be{\begin{equation}}
\def\ee{\end{equation}}
\def\ba{\begin{eqnarray}}
\def\ea{\end{eqnarray}}
\def\x{\bf x}
\def\B{\rm B}
\def\D{\rm D}
\def\E{\rm E}
\def\F{\rm F}
\def\G{\rm G}
\def\H{\rm H}
\def\I{\rm I}
\def\J{\rm J}
\def\K{\rm K}
\def\L{\rm L}
\def\M{\rm M}
\def\P{\rm P}
\def\S{\rm S}
\def\T{\rm T}
\def\V{\rm V}
\title{\Large   Coupled-Channel and Screening Effects in Charmonium Spectrum}
\author{
Bai-Qing Li$^{a,b}$, Ce Meng$^a$ and Kuang-Ta Chao$^a$}

\affiliation{$^a$Department of Physics and State Key Laboratory of
Nuclear Physics and Technology, Peking University, Beijing 100871,
China;
$^b$Department of Physics, Huzhou Teachers College, Huzhou
313000, China}
\date{\today}
\begin{abstract}
Using the same quenched limit as input, we compare the charmonium
spectra predicted by the coupled-channel model and the screened
potential model in the mass region below 4 GeV, in which the
contributions from decay channels involving P-wave (as well as even
higher excited) D mesons can be neglected. We find that the two
models  have similar global features in describing charmonium
spectrum since they embody the same effect of the vacuum
polarization of dynamical light quark pairs. This agreement is
encouraging for using the screened potential to further describe the
mass spectrum of higher charmonium in the region above 4 GeV.
Applying both models will be helpful to clarify the nature of the
newly discovered $"X,Y,Z"$ mesons in B-factories and other
experiments. In particular, we show the S-wave decay coupling effect
on lowering the $\chi_{c1}(2P)$ mass towards the $D\bar {D^*}$
threshold, in support of the assignment of the X(3872) as a
$\chi_{c1}(2P)$-dominated charmonium state.
\end{abstract}
 \maketitle
\section{Introduction}

Studies on heavy quarkonium spectroscopy have been stimulated
greatly in recent years by the discovery of many hidden charm
states, the so-called $"X,Y,Z"$ mesons~\cite{Olsen-XYZ} in
B-factories and other experiments. The QCD-inspired interquark
potential models, such as the Cornell model~\cite{Eichten:1978tg},
are successful in predictions of charmonium and bottomonium spectra
below the open flavor thresholds. However, the existence of open
charm thresholds can change the charmonium spectrum significantly
through the virtual charm meson loops. These coupled-channel effects
were considered also in the Cornell model~\cite{Eichten:1978tg}, and
techniques were further developed by the unitaritized quark
model~\cite{Tornqvist84:CCM} based on the $^3P_0$ quark pair
creation model~\cite{Le Yaouanc:1972ae}. Along this line and with
the updated parameters, the coupled-channel effects in the
charmonium spectrum have been further studied during these
years~\cite{Barnes07:CCM,Pennington07:CCM,Kalashnikova:2005ui}.
These studies provide important information on the identifications
of the $"X,Y,Z"$ mesons.

The quark potential model is subject to modification due to quantum
fluctuation, i.e., the creation of light quark pairs, which may be
compensated by the virtual hadron loops in the coupled-channel
model. For this reason, the quark potential
model~\cite{Eichten:1978tg}, which incorporates a Coulomb term at
short distances and a linear confining term at large distances, may
be called as the quenched potential model. On the other hand, the
vacuum polarization effect of the dynamical fermions may soften the
linear potential at long distances~\cite{Laermann:1986pu}, and cause
the screening effect, which may be discussed phenomenologically as
the screened potential
model~\cite{Chao:1992et,Ding:1993uy,Ding:1995he}. Such screening or
string breaking effects have been demonstrated, although indirectly,
by the simulations of unquenched lattice QCD~\cite{Bali:2005fu}.
This effect is also implied by calculations within some holographic
QCD models \cite{Armoni08:Screening}. The screened potential model
has been used to reexamine the charmonium
spectrum~\cite{Li09:ScreenedPotential} recently, and it is found
that the masses of higher charmonium states are lowered, compared to
the quenched potential model~\cite{Barnes2005}, and the mass
suppression tends to be strengthened  from lower levels to higher
ones. Such tendency can also be found in the calculations of the
coupled-channels model, such as those in
Ref.~\cite{Pennington07:CCM}. It is not very surprising since,  in
the parton-hadron duality picture, the two models embody the same
effects of light quark pairs.

Comparing to the screened potential model, the coupled-channel model
is more difficult to handle in practice, especially in the case when
the P-wave (\,and higher excited\,) $D$ and $D_{s}$ mesons as the
intermediate states are involved. However, the latter can describe
the near-threshold
effect~\cite{Pennington07:CCM,Kalashnikova:2005ui,Rosner-Bugg:threshold-effects},
which has been ignored in the former. It is then interesting to
compare the two models using the same quenched limit as input in the
domain of charmonium spectrum. The comparison has twofold meaning:
the coupled-channel model can be helpful to establish the form of
the screened potential and to determine the screening parameter
$\mu$ in the mass region below 4 GeV, where the P-wave $D$ mesons
are expected to be decoupled; whereas the screened potential model
can be helpful to normalize the global features of the coupled
channel model in the mass region above 4 GeV.

In this paper, we will compare these two models in the mass region
below 4 GeV for charmonium spectrum with the same quenched limit. We
will introduce the two models in turn. And we will compare results
of these two models numerically, and finally a summary will be
given.

\section{Quenched and screened potential models}

We choose the Cornell model~\cite{Eichten:1978tg} as the quenched
limit, in which the potential has the well known form:
\be V(r)=-\frac{4}{3} \frac{\alpha_c}{r}+\lambda r +C,
\label{V:Cornell}\ee
where the first term denotes the color Coulomb force in the
one-gluon exchange approximation due to asymptotic freedom of QCD at
short distances, and the second term denotes the linear confining
potential, which is consistent with both the rotating string
picture~\cite{Goddard73:string} and the quenched lattice
calculations (see~\cite{Bali01:report} for a review and references).
The constant $C$ in (\ref{V:Cornell}) is the renormolization term.

In principle, one can relate the parameter $\alpha_c$ to the running
coupling constant $\alpha_s(m_c v)$ in QCD, where $v^2\approx 0.3$
is the charm quark velocity squared in the charmonium rest frame,
and relate $\lambda$ to the string tension
$T=1/(2\pi\alpha')\sim0.18$ GeV$^2$, where $\alpha'$ denotes the
Regge slope in the rotating string picture~\cite{Goddard73:string}.
Thus, we choose
\be
\alpha_c=0.55,~~~\lambda=0.175~\mbox{GeV}^2\label{Parameter:Cornell}\ee
in the following analysis.

To restore the hyperfine and fine structures of the charmonium
spectrum, one needs to introduce the spin-dependent potential
$V_{sd}$, which is relativistically suppressed compared to $V$ in
(\ref{V:Cornell}). Assuming the Lorentz structure of the linear
confining force is of scalar type, the spin-dependent potential can
be derived from (\ref{V:Cornell}) by the standard Breit-Fermi
expansion to order $v^2/c^2$, and has the form~\cite{Barnes2005}
\begin{eqnarray}
\label{V:sd}
 V_{sd}(r) &=&(\frac{2\alpha_c}{m_c^2r^3}-\frac{\lambda}{2m_c^2r})\vec {\L} \cdot \vec {\S}+\frac{32\pi \alpha_c}{9m_c^2}\tilde \delta(r)
\vec {\S}_c \cdot \vec {\S}_{\bar c} \nonumber\\
&&+\frac{4\alpha_c}{m_c^2r^3}(\vec {\S}_c \cdot \vec {\S}_{\bar
c}+\frac{3(\vec {\S} \cdot \vec {r})(\vec {\S} \cdot \vec
{r})}{r^2}).
\end{eqnarray}
where $\vec {\S}=\vec {\S}_c+\vec {\S}_{\bar c}$ is the total spin,
$m_c$ the charm quark mass, and the smeared delta function is taken
to be $ \tilde \delta(r)=(\sigma/\sqrt{\pi})^3\, e^{-\sigma^2 r^2}$
with $\sigma=1.45$ GeV~\cite{Barnes2005}. Then, the Hamiltonian for
the quenched potential model is given by
\be H_0 = 2m_c+\frac {{\vec{P}}^2}{m_c}+V(r)+V_{sd}(r),
\label{H0:charmonium}\ee
where the kinematic energy term have been included explicitly.

As mentioned above, the linear confining potential will be softened
by the vacuum polarization induced by the dynamical quark pair
creation. Such unquenched effect can be roughly accounted for by
modifying the long distant behavior of $V(r)$ in (\ref{V:Cornell}).
Following Refs.~\cite{Ding:1993uy,Ding:1995he}, we use the screened
potential
\begin{equation}
\label{V:screened}
 V_{scr}(r)=-\frac{4}{3}\frac{\alpha_c}{r}+\lambda r \frac
{1-e^{-\mu r}}{\mu r}+C
\end{equation}
to substitute $V(r)$. The first term on the right-hand side of
(\ref{V:screened}) is taken to be the same as that in
(\ref{V:Cornell}) due to its short distance nature. The screening
parameter $\mu$ sets the scale of distance at which the string
breaks. We choose
\begin{equation}
\label{mu:screening}
 \mu=0.075~\mbox{GeV},
\end{equation}
of which the inverse is about two times of the radius of $D$ meson.
Needless to say, the right hand side of (\ref{V:sd}) should also be
modified accordingly.
\begin{figure}[t]
\begin{center}
\vspace{0cm} \hspace*{0cm}
\scalebox{0.5}{\includegraphics[width=16cm,height=9cm]{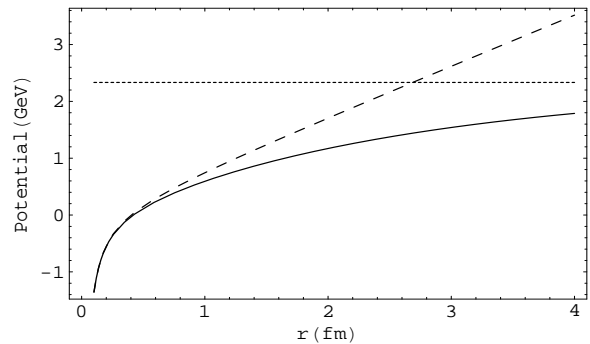}}
\end{center}
\vspace{0cm}\caption{Comparison of the Cornell potential $V(r)$
(dashed line) and the screened potential $V_{scr}$ (solid line) with
$r=0.1\mbox{-}4$ fm. The constant terms are neglected and asymptotic
limit of $V_{scr}$ is shown in the dotted
line.}\label{Fig:Potential}
\end{figure}

At short distances, the screened potential $V_{scr}$ in
(\ref{V:screened}) has the limit consistent with that of $V(r)$ in
(\ref{V:Cornell}). However, the difference between the potentials
increases when $r$ increases, and finally $V_{scr}$ goes to a
constant $\lambda/\mu+C$ when $r\to\infty$, as shown in
Fig.~\ref{Fig:Potential}. As a result, the charmonium spectrum in
the screened potential model will be suppressed compared to that in
the quenched potential model, especially for the higher exited
states. Furthermore, in the screened charmonium spectrum, there will
be a saturation energy of about 5-6 GeV, at which the $c\bar{c}$
quark pair can not be bound together at all.

Let's stress again that the parameters in (\ref{Parameter:Cornell})
and (\ref{mu:screening}) may not be the same as those chosen
in~\cite{Li09:ScreenedPotential}, since our purpose here is to
compare the spectra in the two models rather than to get a good fit
of the spectrum to the experimental data.

\section{The $^3P_0$-based non-relativistic coupled channel model }

The light quark pair creation from vacuum is assumed to share the
same quantum number, $0^{++}$, as the vacuum in the $^3P_0$
model~\cite{Le Yaouanc:1972ae}. In the non-relativistic limit, the
$^3P_0$ model can be represented by the
Hamiltonian~\cite{Ackleh96:3P0}
\begin{eqnarray} H_{QPC}=2m_q \gamma \int d^{3}
\label{Hqpc} \vec{x}\bar{\Psi}_q\Psi_q,
\end{eqnarray}
where $m_q$ is the mass of the produced quark, and $\gamma$ is a
constant reflecting quark pair creation strength, which can be
determined phenomenologically by the OZI-allowed decay widths of
charmonia~\cite{Barnes2005}. There are no fundamental reasons to
enhance the heavier quark creation relative to lighter ones, which
is implied by the factor of $2m_q$ in (\ref{Hqpc}). Therefore,
following Ref.~\cite{Kalashnikova:2005ui}, we use the effective
strength $\gamma_s=\frac{m_q}{m_s}\gamma$ for the strange quark,
where $\frac{m_q}{m_s}$ is the ratio of constituent masses of up and
down quarks($m_q=m_u=m_d$) and strange quark($m_s$). Following
Ref.~\cite{Kalashnikova:2005ui}, we choose $\gamma=0.322$.

Light quark pair creation can result in mixing between the bare
charmonium state $\psi_0$($c\bar c$) and the open charmed meson pair
$B$($c \bar q$) and $C$($q \bar c$). Thus, neglecting the mixing
among the bare states~\cite{Barnes07:CCM}, the physical state $\Psi$
can be represented by
\begin{eqnarray}
 \mid \Psi \rangle =a_0 \mid \psi_0 \rangle + \sum_{BC}\int d
 \nu~ c_{BC}(\nu)  \mid   BC; \nu \rangle,\label{Psi:CCM}
\end{eqnarray}
where $\nu$ denotes the variable of three-momenta of the $BC$
system. The coefficients $a_0$ and $c_{BC}(\nu)$ are understood to
be subject to the normalization of the corresponding wave functions.

The Hamiltonian of this coupled-channel system can be formally
written as
\begin{eqnarray}
 H=H_0+H_{BC}+H_{QPC}.\label{H:CCM}
\end{eqnarray}
Here, $H_0$ has been introduced in (\ref{H0:charmonium}) but can be
different from the one in the quenched potential model by different
renormalization constant $C$. The Hamiltonians $H_0$ and $H_{BC}$
can only act on the states $\psi_0$ and $BC$, respectively, and give
the bare spectra of them:
\begin{eqnarray}
\label{Hoa}
 H_0 \mid  \psi_0 \rangle &=& M_0 \mid
\psi_0 \rangle,\label{M0:defination}\\
H_{BC} \mid BC \rangle &=& E_{BC}(\nu) \mid BC; \nu
\rangle,\label{M_BC:defination}
\end{eqnarray}
where $M_0$ is the so-called bare mass of the bare charmonium state
and
\be E_{BC}(\vec{P}_B,\vec{P}_C)=\sqrt{M_B^2+\vec{P}_B^2}+
\sqrt{M_C^2+\vec{P}_C^2}, \ee
provided that the inner interactions between $B$ and $C$ can be
neglected. On the other hand, the $H_{QPC}$ in (\ref{H:CCM}), which
is defined in (\ref{Hqpc}), can acts only between $\mid \psi_0
\rangle$ and $\mid  BC; \nu \rangle$.

The Hamiltonian $H$ in (\ref{H:CCM}) defines the physical mass $M$
of the state $\Psi$ as
\begin{eqnarray}\label{M:defination}
 H \mid \Psi  \rangle = M \mid \Psi  \rangle,
\end{eqnarray}
and the mass $M$ can be obtained by solving the multi-channel
Shr\"{o}dinger equation (\ref{M:defination}). Substituting
(\ref{Psi:CCM})-(\ref{M_BC:defination}) into (\ref{M:defination}),
one can get the integral equation
\be\label{MassEquation:CCM}
 M+\Pi(M) -M_0=0,
\ee
where the self-energy function $\Pi(M)$ is given by
\be\label{Pi:CCM}
 \Pi(M) =\sum_{BC}\int d \nu ~\frac {|\langle BC;\nu \mid H_{QPC} \mid
 \psi_0
 \rangle|^2}
 {E_{BC}(\nu)-M+i \epsilon}.
\ee
When $\mbox{Re}[M]>M_A+M_B$, it is obvious that the function
$\Pi(M)$ will not be real, and the imaginary part is proportional to
the decay width of $\Psi\to BC$. Therefore, one can solve
Eq.(\ref{MassEquation:CCM}) in the complex plane and get the pole
mass $\mbox{Re}[M]$~\cite{Pennington07:CCM}. However, we will define
the coupled-channel mass $M_{cou}$ as~\cite{Kalashnikova:2005ui}
\be\label{M_cou:defination}
 M_{cou}+\mbox{Re}[\Pi(M_{cou})] -M_0=0,
\ee
which is also called the Breit-Wigner mass by the authors of
Ref.~\cite{Pennington07:CCM}.

It is worth emphasizing here the difference between the definitions
of the bare mass in~\cite{Pennington07:CCM} and of ours. The authors
of Ref.~\cite{Pennington07:CCM} once subtract the dispersion
integral in (\ref{Pi:CCM}) at $M_\psi$, and absorb the term
$\Pi(M_\psi)$ into the bare mass definition:
\be\label{M_0:P&W}
 M_0'=M_0-\Pi(M_\psi),
\ee
where $\Pi(M_\psi)$ is real and positive, which can be seen directly
from the definition of the function $\Pi(M)$ in (\ref{Pi:CCM}). If
the subtracted constant $\Pi_n(M_\psi)$ are the same for all the
charmonium states $n$, the bare mass $M_0'$ in (\ref{M_0:P&W}) is
just a rescaling one of $M_0$ and the renormalized mass shift would
not changed. In Ref.~\cite{Pennington07:CCM}, the matrix element
square $|\langle BC\mid H_{QPC} \mid \psi_0 \rangle|^2$ in
(\ref{Pi:CCM}) is simply parameterized by using a exponential form
factor, and the node structure in the wave function of higher
excited state is absolutely neglected. Thus, the subtracted
constants $\Pi_n(M_\psi)$ for these excited states might be
overestimated, and as a result, the renormalized mass shifts in
Ref.~~\cite{Pennington07:CCM} are commonly smaller than those in our
model, as one can seen in the following section.

On the other hand, the wave functions of $\psi_0$, $B$ and $C$ are
needed to determine the matrix element $\langle BC\mid H_{QPC} \mid
\psi_0 \rangle$ in (\ref{Pi:CCM}). These wave functions are usually
chosen as the harmonic oscillator
ones~\cite{Barnes07:CCM,Kalashnikova:2005ui,Ackleh96:3P0}. However,
we determine them by solving the quenched mass equation
Eq.(\ref{M0:defination}) with Eq.(\ref{H0:charmonium}).
But for simplicity, we will neglect the corrections due to the
spin-dependent potential $V_{sd}$ to these wave function.

\section{Numerical Results and Discussion}
\begin{table*}
\caption{Chamonium spectra and mass shifts in different models in
units of MeV. Here, the subscripts $que$, $cou$, and $scr$ denote
the results obtained from the quenched potential model, the coupled
channel model, and the screened potential model, respectively. The
mass shifts $\Delta M_{cou}$ and $\Delta M_{scr}$ are listed in the
5th and 6th columns, respectively. The results of
Ref.~\cite{Pennington07:CCM} are also listed. The bare mass in the
7th column is copied from Ref.~\cite{Barnes2005}. All the quantities
listed here should be understood as the renormalized ones.}
\begin{center}
\begin{tabular}{|c|c|c|c|c|c|c|c|c|c|}
 \hline
 &\multicolumn{5}{c}{Our results}&\multicolumn{3}{|c|}{Results of Ref.~\cite{Pennington07:CCM}}\\
 \cline{2-9}
 ~~states~~ &~~~~$M_{que}$~~~~ &~~~~$M_{cou}$~~~~ &~~~~$M_{scr}$~~~~&~~~$\Delta M_{cou}$~~~&~~~$\Delta M_{scr}$~~~&~~~~$M_0'$~~~~&~~~~$M_{cou}'$~~~~&~~~$\Delta M_{cou}'$~~~\\
 \hline
 $1{}^1S_{0}$ &2980&2980 &2980.0&0&0&2982&2982&0\\
 \hline
 $1{}^3S_{1}$ &3112&3100 &3105&-12&-7&3090&3090&0\\

\hline
 $1{}^1P_{1}$ &3583&3531 &3539&-52&-44&3516&3514&-2\\
\hline
 $1{}^3P_{0}$ &3476&3441 &3448&-35&-28&3424&3415&-9\\
\hline
 $1{}^3P_{1}$ &3568&3520 &3526&-48&-42&3505&3489&-16\\
\hline
 $1{}^3P_{2}$ &3628&3565 &3577&-63&-51&3556&3550&-6\\

\hline
 $2{}^1S_{0}$ &3697&3635 &3626&-62&-71&3630&3620&-10\\
 \hline
 $2{}^3S_{1}$ &3754&3674 &3674&-80&-80&3672&3663&-9\\

\hline
 $1{}^1D_{2}$ &3895&3818 &3805&-77&-90&3799&&\\
\hline
 $1{}^3D_{1}$ &3878&3794 &3790&-84&-88&3785&3745&-40\\
\hline
 $1{}^3D_{2}$ &3896&3818 &3805&-78&-91&3800&&\\
\hline
 $1{}^3D_{3}$ &3903&3823 &3812&-80&-91&3806&&\\

\hline
 $2{}^1P_{1}$ &4042&3961 &3909&-81&-133&3934&3929&-5\\
\hline
 $2{}^3P_{0}$ &3948&3915 &3839&-33&-109&3852&3782&-70\\
\hline
 $2{}^3P_{1}$ &4030&3875 &3900&-155&-130&3925&3859&-66\\
\hline
 $2{}^3P_{2}$ &4085&3966 &3941&-119&-144&3972&3917&-55\\
\hline
\end{tabular}
\end{center}
\label{Tab:total-results}
\end{table*}

As mentioned in Sec. I, our aim is to compare the charmonium spectra
or the renormalized mass shifts
\be\label{massshifts:defination}
 \Delta M_{cou}=M_{cou}-M_{que},~~~\Delta M_{scr}=M_{scr}-M_{que},
\ee
where the subscripts $que$, $cou$, and $scr$ denote the results
obtained from the quenched potential model, the coupled-channels
model, and the screened potential model, respectively. The quenched
mass $M_{que}$ can be related to the bare mass $M_0$ in the
coupled-channels model by the relation:
\be\label{M_que:defination}
 M_{que}-C_{que}=M_{0}-C_{cou},
\ee
where $C_{que}$ and $C_{cou}$ denote the renormalization constants
in the quenched potential model and the coupled-channels model,
respectively. Thus from Eq.~(\ref{M_cou:defination}), the mass shift
$\Delta M_{cou}$ can be given by
\be\label{DelM_cou:relation}
 \Delta M_{cou}=-\mbox{Re}[\Pi(M_{cou})]-C_{que}+C_{cou}.
\ee

To improve the reliability of the calculation of the coupled-channel
model, we restrict the mass region of charmonium spectrum  to be
below 4 GeV, in which the decays to P-wave (\,and higher excited\,)
$D$ and $D_{s}$ mesons are kinematically forbidden, and only the
S-wave $D$ and $D_{s}$ mesons are involved. So in the following we
use the S-wave $D_{(s)}$ mesons, of which the masses are well
determined and the widths can be neglected~\cite{PDG08}, as the
intermediate states only.

In the mass region below 4 GeV, the charmonium spectrum consists of
the 1S, 2S, 1P, 2P and 1D levels. For convenience, we choose the
renormalization condition such that the predicted $\eta_c$ masses in
the three models are fixed to be the measured value~\cite{PDG08},
i.e., 2980 MeV. Together with the inputs of the quark masses
\be\label{Parameter:mq}
m_c=1.7~\mbox{GeV},m_q=0.33~\mbox{GeV},m_s=0.5~\mbox{GeV}, \ee
the renormalization condition determines the constant terms in
(\ref{V:Cornell}):
\be\label{Parameter:mq}
C_{que}=-419~\mbox{MeV},C_{cou}=-272~\mbox{MeV},C_{scr}=-403~\mbox{MeV}.
\ee

We list the numerical results of the mass spectra and the mass
shifts in Tab.~\ref{Tab:total-results}. The subscripts $que$, $cou$,
and $scr$ denote the results obtained from the quenched potential
model, the coupled channel model, and the screened potential model,
respectively. And the mass shifts have been defined in
(\ref{massshifts:defination}). In Tab.~\ref{Tab:total-results}, both
$\Delta M_{cou}$ and $\Delta M_{scr}$ have minus sign, and on the
whole, they are consistent with each other. This can also be seen
from Fig. 2.

In addition, for the 1S, 2S, 1P and 1D levels, which either couple
the $D$ meson pair in P-wave or lie far away from the threshold of
the $D$ meson pair, the mass shifts $\Delta M_{cou}$'s are
comparable  to each other, which is consistent with the first Hadron
Loop Theorem derived by the authors of Ref.~\cite{Barnes07:CCM} in
the approximation of equal masses for charmed mesons.

However, if the $c\bar c$ pair couples to $D$  meson pair in S-wave
and the mass $M$ calculated in the coupled channel model (namely
$M_{cou}$) in (\ref{Pi:CCM}) is close to the threshold $M_B+M_C$,
the self-energy $\Pi$, then the mass shift $\Delta M_{cou}$, will
strongly depend on $M(M_{cou})$. This is well known as the S-wave
threshold effect~\cite{Rosner-Bugg:threshold-effects}, and as a
result, the loop theorem~\cite{Barnes07:CCM} will be violated.

This is just the case for the 2P charmonium states, since for some
of them the  coupled-channel masses  $M_{cou}$ (see
Tab.~\ref{Tab:total-results}) are close to the thresholds of the
S-wave channels $D\bar{D}^*+c.c.$ and $D^*\bar{D}^*$\footnote{The
contributions to the mass shifts from the $D_s^{(*)}\bar D_s^{(*)}$
channels are very small due to the small strange quark pair creation
strength $\gamma_s$.}. More in details, for the 2$^3P_0$ state
($\chi_{c0}'$), the coupled-channel mass $M_{cou}(\chi_{c0}')=3915$
MeV, which is fairy far from the threshold of the S-wave channels
$D\bar{D}$ and $D^*\bar{D}^*$. Consequently, the mass shift is as
small as 33 MeV. As a second example, the coupled-channel masses of
the 2$^1P_1$ ($h_c'$) and 2$^3P_2$ ($\chi_{c2}'$) states are roughly
equal. However, their mass shifts induced by the $D^*\bar{D}^*$, of
which the threshold is closest to their coupled-channel masses, are
different by a factor of 2 (see Tab. I in Ref.~\cite{Barnes07:CCM}).
As a result, the mass shift of 2$^1P_1$ state is smaller than that
of 2$^3P_2$ state. Finally, the coupled-channel effect of 2$^3P_1$
($\chi_{c1}'$) state should be most significant since the mass
$M_{cou}(\chi_{c1}')=3875$ MeV is very close to the threshold of
$D^0\bar{D}^{*0}/D^+\bar{D}^{*-}+c.c.$. This result can also give
support to the $\chi_{c1}'$ assignment of
$X(3872)$~\cite{X3872:charmonium}.
\begin{figure}
\label{fig2}
\includegraphics[width=8cm,height=7cm]{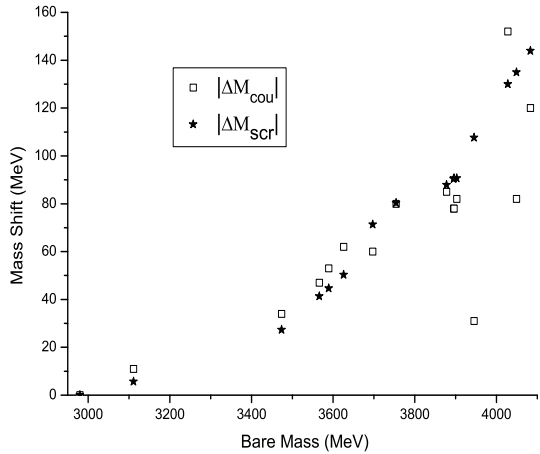}
\caption{Mass shifts $\Delta M_{cou}$ and $\Delta M_{scr}$ varying
with bare mass $M_{0}$.}
\end{figure}

It needs emphasizing here that the closeness of
$M_{cou}(\chi_{c1}')$ to the threshold of $D^0\bar{D}^{*0}$ is not
very sensitive to the bare mass of $\chi_{c1}'$. This can be seen
from Fig.~3, where the physical mass $M_{cou}$ dependence of the
unrenormalized mass shift $-\mbox{Re}[\Pi(M_{cou})]$ for the
$2^3P_1$ state is shown. The relation between the unrenormalized
mass shift and the renormalized one $\Delta M_{cou}$ is given in
(\ref{DelM_cou:relation}). From Fig.~3 one can see the mass shift
function is strongly dependent on the physical mass. As a result,
the slope of the mass shift curve is very large near the threshold
and "attract" the mass $M_{cou}(\chi_{c1}')$ towards the threshold.
That means, if one changes the bare mass, say, by 50 MeV, the change
of the mass $M_{cou}(\chi_{c1}')$ is only about 10-15 MeV as can be
roughly found from the figure. This is just a realization of the
S-wave threshold effect~\cite{Rosner-Bugg:threshold-effects} in our
coupled-channels model. More precisely, the curve in Fig. 3 shows
the cusps in the neutral and charged $D^*\bar D$ channels
numerically (see the second paper
in~\cite{Rosner-Bugg:threshold-effects} for more discussions). Thus,
the physical mass of $\chi_{c1}'$ is quite natural to be close to
the threshold of $D^0\bar{D}^{*0}$ and thus $\chi_{c1}'$ may be a
good candidate for the $X(3872)$~\cite{X3872:charmonium}.
\begin{figure}
\label{Fig:23P1self_energy}
\includegraphics[width=8cm,height=7cm]{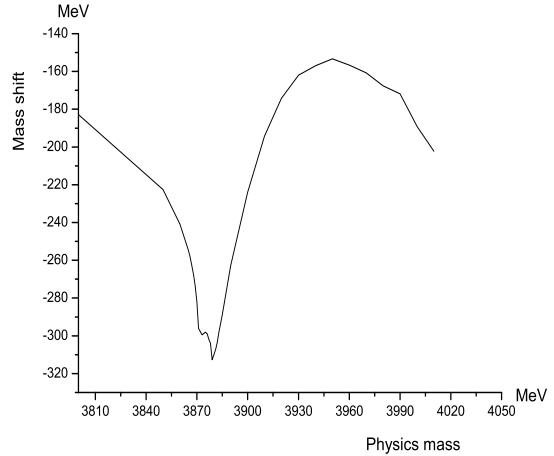}
\caption{The physical mass $M_{cou}$ dependence of the
unrenormalized mass shifts $-\mbox{Re}[\Pi(M_{cou})]$ for the
$2^3P_1$ state.}
\end{figure}

We also list the results of Ref.~\cite{Pennington07:CCM} in
Tab.~\ref{Tab:total-results} for comparison.  As we have mentioned,
the definition of the renormalized bare mass $M_0'$ (\ref{M_0:P&W})
is a little different from that of $M_0$. Moreover, the node effect
in the wavefunction overlap integral for a higher excited charmonium
decaying into charmed mesons is not considered in
Ref.~\cite{Pennington07:CCM}. As a result, the renomalized mass
shift $\Delta M_{cou}'=\Pi(M_\psi)-\mbox{Re}\Pi(M)$
in~\cite{Pennington07:CCM} tends to be smaller than the one in
(\ref{DelM_cou:relation}) for excited states as has been analyzed in
the last section. Furthermore, the authors of
Ref.~\cite{Pennington07:CCM} use the available results of the
quenched potential model~\cite{Barnes2005}, in which different
parameters from ours are chosen, as their bare mass inputs (the 7th
column in Tab.~\ref{Tab:total-results}).
However, comparing the mass shifts in the 5th and 9th columns in
Tab.~\ref{Tab:total-results}, one can find that they both have
similar features as that of $\Delta M_{scr}$. This indicates that
the screened potential in (\ref{V:screened}) depicts the main
feature of the vacuum polarization effect of the dynamical quark
pair creation, although it fails to describe some fine structures,
such as those induced by the near S-wave threshold effects.
\begin{figure}
\label{Fig:massshifts-cog}
\includegraphics[width=8cm,height=7cm]{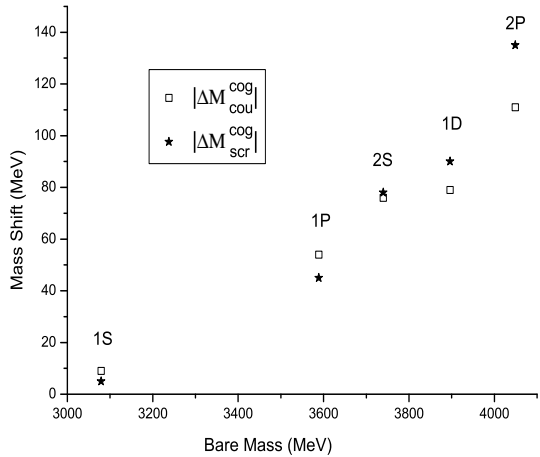}
\caption{The trajectories of the c.o.g.~shifts $\Delta
M^{cog}_{cou}$ and $\Delta M^{cog}_{scr}$ varying with the bare mass
$M_{0}$.}
\end{figure}

To compare the global features of spectra for the coupled-channel
model and the screened potential model more directly, we also
illustrate the c.o.g. shifts $\Delta M^{cog}_{cou}$ and $\Delta
M^{cog}_{scr}$ in Fig.~4. Here, the c.o.g.~is defined as
\be\label{cog:defination}
 M^{cog} =\frac{M_{sig}+3M_{tri}^{cog}}{4},\ee
where $M_{sig}$ denotes the mass of the spin-singlet state, and
$M_{tri}^{cog}$ the c.o.g.~of the spin-triplet states. From Fig.~4,
one can see both the mass shift trajectories exhibit good behaviors,
and they are roughly consistent with each other as the above
analysis. However, the increase of the mass shift $\Delta
M^{cog}_{scr}$ tends to be faster than that of $\Delta
M^{cog}_{cou}$ in the higher mass region. This seems to indicate
that the potential in (\ref{V:screened}) somewhat overestimates the
screening effect. But it is not the whole story since the P-wave
(\,and higher excited\,) $D_{(s)}$ mesons contributions, which have
been neglected, also tends to enhance the mass shift $\Delta
M^{cog}_{cou}$ in the same higher mass region.

\section{Summary}

We compare the charmonium spectra predicted by the coupled-channel
model and the screened potential model in the mass region below 4
GeV, in which the contributions from  channels involving the P-wave
(\,and higher excited\,)$D$ and  $D_{s}$ mesons can be neglected. We
use the same quenched limit for the two models. And for the
coupled-channel model, we use the wave functions obtained by the
quenched potential model in the non-relativistic limit to determine
the hadronic transition matrix elements (wavefunction overlap
integrals) in (\ref{Pi:CCM}).

We find that the two models have similar global features in
describing the charmonium spectrum as expected since they embody the
same effect of the vacuum polarization of the dynamical quark pair
creation. This agreement is encouraging for using the screened
potential to further describe the mass spectrum of higher charmonium
in the region above 4 GeV. Applications of both models will be
helpful to clarify the nature of the newly discovered $"X,Y,Z"$
mesons~\cite{Olsen-XYZ}. In particular, we show the S-wave coupling
effect on lowering the $\chi_{c1}(2P)$ mass towards the $D\bar
{D^*}$ threshold as support to assign the X(3872) as a
$\chi_{c1}(2P)$-dominated charmonium state.

\begin{acknowledgments}
We thank D. Bugg for useful communications. This work was supported
in part by the National Natural Science Foundation of China (No
10675003, No 10721063), and by China Postdoctoral Science Foundation
(No 20080430263).
\end{acknowledgments}

\end{document}